\documentclass{aa}  
\nolinenumbers

\usepackage{graphicx}
\usepackage{txfonts}
\usepackage{soul}
\usepackage{graphicx}	
\usepackage{amsmath}	
\usepackage[colorlinks,citecolor={blue}]{hyperref}
\usepackage{bbm}
\usepackage{cleveref}
\usepackage{multirow}
\usepackage{pgfplots}
\usepackage{placeins}
\usepackage{tikz}
\usepackage{upgreek}
\usepackage{booktabs}
\definecolor{ccqqqq}{rgb}{0.8,0,0}
\graphicspath{./Figures/}

\pgfplotsset{compat=1.15}

\crefname{equation}{equation}{equations}
\Crefname{equation}{Equation}{Equations}
\crefname{figure}{Fig.}{Figs.}
\Crefname{figure}{Fig.}{Figs.}
\crefname{table}{Table}{Tables}
\Crefname{table}{Table}{Tables}
\crefname{section}{Section}{Sections}
\Crefname{section}{Section}{Sections}

\begin{document}

   \title{Cluster connectivity in DES Y3 data: impact on galaxy properties}

   \subtitle{}

   \author{ C.~Gouin\inst{1}, C.~Laigle\inst{1}, J.~Prat\inst{2,3}, L.~O.V.~Edwards\inst{4},   P.~Ganeshaiah~Veena\inst{5},  D.~Galárraga-Espinosa\inst{6, 7}, G. Gopidas\inst{4} 
   }

   \institute{ CNRS and Sorbonne Universit\'e, UMR 7095, Institut d'Astrophysique de Paris, 98 bis, Boulevard Arago, F-75014 Paris, France
            \and 
            Departament de Fìsica, EEBE, Universitat Politècnica de Catalunya, c/ Eduard Maristany 16, E-08019 Barcelona, Spain
            \and
            Nordita, Stockholm University and KTH Royal Institute of Technology, Hannes Alfv\'ens v\"ag 12, SE-10691 Stockholm, Sweden
            \and  California Polytechnic State University, 1 Grand Ave. San Luis Obispo, CA, 93407, USA
             \and  Department of Physics, University of Genova, Via Dodecaneso, 33, Genoa, Italy
             \and
             Kavli IPMU (WPI), UTIAS, The University of Tokyo, Kashiwa, Chiba 277-8583, Japan
             \and
             Center for Data-Driven Discovery, Kavli IPMU (WPI), UTIAS, The University of Tokyo, Kashiwa, Chiba 277-8583, Japan
             }

   \date{Received ; submitted}

 
  \abstract{
   {{Connectivity, namely the number of cosmic filaments reaching out of groups or clusters, has recently emerged as an informative way to quantify environment. With the advent of wide area photometry surveys, it becomes possible to measure connectivity on a statistically significant sample of clusters. }}
   {{In this work, we extract the cosmic web from the DES Y3 photometric data and we measure the connectivity around clusters from several public catalogs up to $z<1$.}}
   {{Cosmic filaments are extracted in 2D with {\sc DisPerSE} (the parametrization of which is calibrated on simulations) in tomographic slices of fixed comoving thickness of 220~Mpc, with a 
   constant sampling across the redshift range ($\sim 0.1$~galaxies ${\rm Mpc}^{-2}$).}}
   We first recover the well-established trend that cluster connectivity increases with cluster mass at all redshifts. We then investigate the properties of cluster member galaxies and find that, at fixed mass, clusters with a higher fraction of red (elliptical) galaxies are more strongly connected to the cosmic web. Furthermore, by examining the radial profiles of the red and elliptical galaxy fractions, we obtain hints that galaxy populations may undergo pre-processing within filamentary structures before being accreted into clusters.
{{Overall, we demonstrate that in despite of the photometric redshift uncertainties and the projection effects, our method allows for the quantification of the cosmic web environment around clusters in broad-band photometric surveys.  This work is consistent with results based on the  \textit{Euclid} Quick release 1 (Q1). We plan to publicly release the cluster connectivity catalog as well as the cosmic filament catalogue.}}}

\titlerunning{Connectivity in DES Y3 data}
\authorrunning{C. Gouin et al.}
   \maketitle
%


\section{Introduction}


The anisotropic large-scale distribution of matter in the Universe forms a complex web-like structure on multiple spatial scales,  known as the \textit{cosmic web} \citep{deLapparent86, Bond1996}. Galaxy clusters and super-clusters are connected to cosmic filaments, which are themselves found at the intersection of walls surrounding voids. 
The formation of the cosmic-web is well understood in the context of the hierarchical growth of structure \citep{Zeldovich1982}. 
Anisotropic matter fluctuations in the early universe are amplified by gravity and result in the complex distribution we see today. It has been shown in numerical simulations that almost $50\%$ of dark matter and $80\%$ of galaxies are in filaments \citep{cen06,aragoncalvo10,cautun_evolutionCosmicWeb_2014, GaneshaiahVeena_cosmicBallet2_2018}. Cosmic filaments are therefore the primary environment in which galaxies grow and evolve. Theoretically, it is expected that the properties of cosmic web filaments shape masses, star-formation rates \citep[e.g.][]{song21,winkel21,daniela23,bulichi2024}, morphologies and alignments of galaxies \citep[e.g.][]{Codis12,GVeena_cosmicBallet2_2019,malavasi22,Zhang23} growing within them, not only through their local density \citep[e.g.][amongst others]{Dressler1980, Lewis2002, Ball2008, Pandey2017,Kuutma2020,okane2024}, but also because the anisotropic geometry of the matter distribution generates tides which in turn impact the accretion onto galaxies \citep[e.g.][]{Musso2018,MonteroDorta2024}.
We have now observational evidence which corroborates these expectations, where galaxy properties vary depending on their location in cosmic filaments \citep[e.g.][]{malavasi2017,Laigle2018,Castignani22,Hoosain2024,Santiago-Bautista20,laigle2026}.

In addition to its impact on galaxies in filaments, the geometry of the large-scale cosmic web environment is also expected to shape groups and clusters. Groups and clusters correspond to peaks of the density fields on which cosmic filaments intersect. Connectivity is the number of filaments to which a given cluster is connected. Recently, connectivity has emerged as a new metric to quantify the structure of the cosmic web environment around peaks \citep[e.g.][]{Einasto2020,Gouin20,Cornwell22}. In particular, several studies have highlighted the correlation between connectivity and the state of relaxation of the clusters, suggesting that more connected clusters are less relaxed, since they are more likely to have undergone recent mergers \citep{darraghford19,kraljic20,gouin21}. Observational evidence also suggests that overall more connected clusters contain more passive and elliptical galaxies \citep{gouin2026}.
Beyond its interest for understanding galaxy populations, connectivity has also been proposed as a cosmological probe that  complements classical statistics \citep[e.g.][]{codis2018,Dome2023,Shim2023,boldrini24}. 

Galaxies are discrete and biased tracers of the underlying dark-matter (DM) distribution, but observationally, they represent an easy way to trace cosmic filaments. 
Ideally, we would rely on spectroscopic surveys because in principle they enable a full 3D reconstruction of the cosmic web \citep[e.g.][]{kraljic2026}. However, they are inevitably limited in volume and sampling. When it comes to measuring the connectivity of galaxy clusters, it is pivotal to work on a large volume of data to recover a statistically significant number of clusters. In addition, these catalogues are generally biased towards certain populations of galaxies (e.g. H$\alpha$-emitters, red galaxies, Lyman-break galaxies, etc.), which hampers the completeness and purity of the cluster identification.  In this case, an alternative method is to work in projection with photometric redshifts and to perform a 2D identification of filaments. This approach was taken, for example, by \cite{Laigle2018} and \cite{sarron19}. 

In this work, we use data from the first three years (Y3) of observations from the Dark Energy Survey \citep[DES][]{DES2016}, which covers $\sim$4200 deg$^2$ up to $z\sim1$. By applying the Discrete Persistent Structure Extractor code ({\sc DisPerSE}, \citealt{Sousbie_Disperse_paper1, Sousbie2011b}) 
algorithm on the dataset, we provide a catalog of filaments that covers the DES sky. 
The methodology developed here will serve as a test bed for upcoming even larger photometric data sets, such as those from the Vera C. Rubin Observatory's Legacy Survey of Space and Time  \citep[LSST,][]{Ivezic2019}, which will cover half of the sky, {Roman's High-Latitude Wide-Area Survey \citep{Roman25}} and the ESA satellite Euclid \citep{euclidoverview}, which will cover 14000~deg$^2$.

The paper is organised as follow.
In Section~\ref{sec:data} we describe the data and simulations that we use to generate and validate the filament catalog. The method for extracting filaments is described in section~\ref{sec:cosmic_web}. Our results are presented and discussed in section~\ref{sec:results}. We conclude in section~\ref{sec:conclusion} and outline future work.



%
%

\section{Data}\label{sec:data}


The Dark Energy Survey is a photometric survey that covers about 5000 deg$^{2}$ of the southern sky to a depth of $r<24$, imaging about 300 million galaxies in 5 broadband filters ($grizY$) up to redshift $z \sim 1.4$ \citep{Flaugher2015,DES2016}. In this work we use the publicly released data\footnote{\url{https://des.ncsa.illinois.edu/releases/y3a2/Y3key-catalogs}} from the first three years of observations (DES Y3).



\subsection{\textsc{MagLim} galaxy catalogue and photometric redshifts}\label{SubSect:Maglim}
In this work, we primarily rely on the \textsc{MagLim} galaxy catalogue, a magnitude-limited sample characterized and described   in \citet{Porredon2022}, which is a subset of the DES Y3 Gold Catalog \citep{Sevilla-Noarbe2021}. The mask covers 4132 deg$^{2}$. This galaxy catalog was used as a lens sample in the DES Y3 3$\times$2pt analysis \citep{desy3-3x2pt} and has been extensively tested. It comes with calibrated photometric redshift estimates. 
The magnitude-limited sample (or \textsc{MagLim})  is selected with a magnitude cut that evolves linearly with the photometric redshift estimate: $i < a z_{\rm phot} + b$. This sample was optimized to maximize cosmological information in the DES Y3 3$\times$2pt cosmological in \citet{Porredon2021} taking into account the trade-off between number density and photometric redshift accuracy. The optimization yielded $a=4.0$ and $b=18$. Effectively this selects brighter galaxies at low redshift while including fainter galaxies as redshift increases.  Additionally,  a lower cut is applied to remove stellar contamination (the objects with magnitudes $i > 17.5$). 

We use the photometric redshift estimates publicly released by the DES Collaboration, hereafter $z_\mathrm{DNF}$. These are obtained with the DNF (Directional Neighbourhood Fitting) photometric redshift code \citep{DeVicente2016}. We estimate the redshift errors as the absolute value of the difference between the $z_\mathrm{DNF}$ and the nearest neighbor spectroscopic redshift for the DNF fit. 




Importantly, for the galaxies that are also present in the \textsc{redMaGiC} sample \citep{Rozo2015}, we replace their $z_\mathrm{DNF}$ and corresponding errors by the values in the \textsc{redMaGiC} catalogue. The latter contains a sample of luminous red galaxies (LRGs) with high quality photometric redshifts, as described in detail and characterized in \citet{Rodriguez-Monroy2022}. The fraction of \textsc{MagLim} galaxies in the \textsc{redMaGiC} catalogue is $\sim 14\%$.

\begin{table}
\centering
\setlength{\tabcolsep}{5pt}
\begin{tabular}{c|c|c|c|c}
Slice &  $z$-range & $N_\text{gal}$ &  $n_\text{gal}$  &  $\sigma_z $ \\
\hline
1 & $0.2000 < z < 0.2562$ &   6.1$\times 10^{5}$ & 0.53 & 0.023 \\
2 & $0.2373 < z < 0.2947$ &    6.7$\times 10^{5}$& 0.44  & 0.025 \\
3 & $0.2754 < z < 0.3341$ &    6.1$\times 10^{5}$ & 0.31 & 0.022 \\
4 & $0.3143 < z < 0.3744$ &   6.1$\times 10^{5}$  &  0.25 & 0.024 \\
5 & $0.3541 < z < 0.4156$ &   6.1$\times 10^{5}$ &  0.17 & 0.035 \\
6 & $0.3949 < z < 0.4579$ &   6.3$\times 10^{5}$ &  0.16 & 0.034 \\
7 & $0.4366 < z < 0.5013$ &   6.7$\times 10^{5}$ &  0.14 & 0.023 \\
8 & $0.4794 < z < 0.5458$ &  6.7$\times 10^{5}$ &  0.12 & 0.017 \\
9 & $0.5234 < z < 0.5916$ &   6.9$\times 10^{5}$ &  0.12 & 0.016 \\
10 & $0.5685 < z < 0.6386$ &   7.7$\times 10^{5}$ &  0.12 & 0.021 \\
11 & $0.6149 < z < 0.6871$ &  8.6$\times 10^{5}$ &  0.11 & 0.025 \\
12 & $0.6627 < z < 0.7370$ &  9.6$\times 10^{5}$  &  0.11 & 0.028 \\
13 & $0.7119 < z < 0.7885$ &    1.1$\times10^{6}$  &  0.12 & 0.029\\
14 & $0.7626 < z < 0.8416$ &   1.2$\times 10^{6}$&   0.11 & 0.029 \\
15 & $0.8149 < z < 0.8965$ &   1.3$\times 10^{6}$ &  0.10 & 0.032 \\

\end{tabular}
\caption{Summary description for each of the tomographic slices on which the cosmic web is reconstructed. $N_\text{gal}$ is the number of galaxies from the \textsc{MagLim} galaxy catalogue, $n_\text{gal}$ is the effective number density in units of gal/Mpc$^{2}$ calculated with an area of 4143 deg$^2$ at the median redshift of the slice, and $\sigma_z$ is the median 1$\sigma$ redshift uncertainties for galaxies in that slice.}
\label{tab:samples}
\end{table}

\subsection{Cluster catalogues}\label{SubSect:clusterCat}

\begin{figure*}
\centering
    \includegraphics[width=0.99\textwidth]{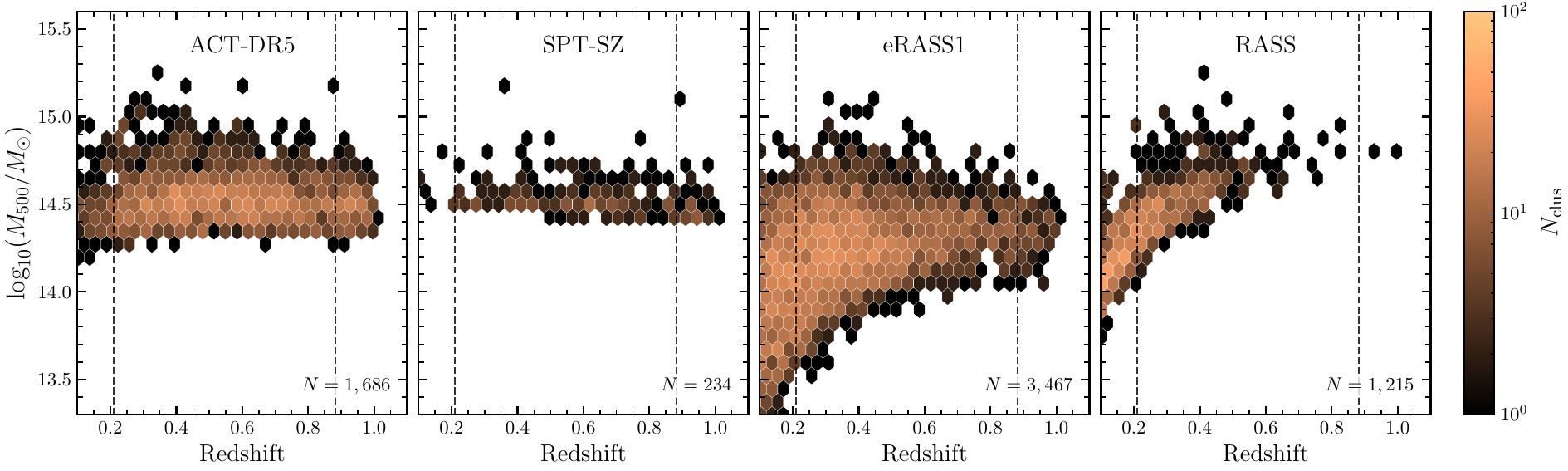} 
    \caption{Redshift and mass distribution of clusters within the DES footprint, for the four external cluster catalogues used in this study. Left: the ACT-DR5 MCMF catalogue \citep[Sunyaev-Zel'dovich effect detection][]{Klein2024ACT}. Middle-left: the SPT-SZ MCMF catalogue \citep[Sunyaev-Zel'dovich effect detection][]{Klein2024SPT}. Middle-right: the eRASS1 catalog \citep[X-Ray detected][]{Bulbul24}. Right: the RASS catalog \citep[X-Ray detected][]{Klein23}. Colour indicates the number of clusters. The two vertical lines indicate the redshift range of our analysis (from slices 1 to 15 according to Tab.~\ref{tab:samples}) 
    }
    \label{fig:clz}
\end{figure*}

Figure~\ref{fig:clz} presents the redshift and mass distributions of the four cluster catalogues within the DES footprint that we used in our analysis. Those catalogues include  clusters detected in X-Ray (eRASS1, RASS) and detected through the Sunyaev-Zel'dovich Effect (namely ACT-DR5 MCMF and SPT-SZ MCMF). The eRASS1 galaxy groups and clusters primary catalogue (left panel) presented in \cite{Bulbul24} has been generated from the SRG/eROSITA All-Sky Survey \citep{Predehl2021} and optical follow-up observations \citep{Kluge2024}. The RASS catalogue (right panel; \citet{Klein23}) is built using MCMF cluster finding algorithm and applied to the second ROSAT All-Sky Survey catalogue \citep{Boller2016_rosat} along with optical data from the 10th data release of the DESI Legacy Imaging Survey \citep{Dey2019}.

The ACT-DR5 MCMF galaxy cluster catalogue (middle-left panel) has been presented in \cite{Klein2024ACT}. Clusters were identified based on the thermal Sunyaev-Zel'dovich Effect (tSZE) in the Atacama Cosmology Telescope data release 5 \citep{naess2020} and confirmed from optical and infrared imaging. The SPT-SZ MCMF catalogue (middle-right panel), presented in \cite{Klein2024SPT}, has been built from detections through the thermal SZE measured on data from the South Pole Telescope \citep{Carlstrom11} and confirmed based on the Dark Energy Survey photometry catalogue. 

In the following, we combined these four catalogues (X-Ray and thermal SZE detections) in the analysis and refer to this compilation as the ``XRAY\_SZ catalogue''. Duplicate clusters are removed, being defined as those present in two or more of the catalogues within 50\arcsec and $z=0.01$ of each other. In addition, clusters closer to 1~deg from the edge of the masked regions are removed from the analysis.



\subsection{Galaxy cluster member properties}

We separate galaxies either based on their morphology, or their colours. 
Galaxy properties are taken from the \citet{cheng2021} galaxy catalog, which is based on the Y3 DES GOLD data \citep{Y3Gold}.  
In their work, they propose a morphological classification based on convolutional neural networks, from five training models based on the $i$-band magnitudes and using visual classification from the Galaxy Zoo 1 catalogue. Each galaxy is assigned an average probability to be spiral and elliptical (pS and pE). We retain in our analysis only galaxies with \texttt{MAG\_AUTO\_I}$<23.0$ and  \texttt{confidence\_flag}$>0$.
As for the colour, the separation between blue and red galaxies is  estimated based on the $r-Y$ apparent color and empirically calibrated on the COSMOS2020 catalogue \citep{Weaver2020}. The details of this calibration are given in Appendix~\ref{app:segreg}.\\
To estimate galaxy membership in each cluster, we compute, for each galaxy, the probability of being at the cluster redshift using the general formalism of \cite{George2011}, which requires both the redshift probability distribution of the cluster and that of the galaxy. We model the cluster redshift distribution as a Gaussian centred on the cluster redshift $z_c$ with a standard deviation of $\sigma_{z_c}=0.01$. Similarly, each galaxy redshift probability distribution is modelled as a Gaussian centred on its redshift $z_g$ with an associated uncertainty $\sigma_{z_g}$ provided by the galaxy catalogue. Following \cite{Castignani2016} and \cite{Sarron2021}, we thus define the membership probability $P_{\rm mem}$ for a galaxy $i$ to belong to a given cluster $C$ as a posterior probability:
\begin{equation}
 P_{{\rm mem},i} \equiv P(i \in C \ | \ {\rm PDF}_{i}(z),{\rm PDF}_{C}(z)) 
\end{equation}
It is obtained by a convolution of the redshift PDFs of the galaxy and the cluster, and this probability is rescaled such that the maximum achievable probability is one.
We refer to \cite{Sarron2021}, and its appendix~C, for detail on this method of probabilistic membership assignment \citep[also used in][]{gouin2026}. \\
For each cluster, we retain as member galaxies only those with $P_{\rm member}>0.5$ and projected distance to the cluster centre of $d_{\rm proj}<2\times R_{\rm 500c}$. When varying these thresholds, the strength of the signal measured in Sec.~\ref{sec:results} slightly changes but the trend remains qualitatively the same.
Moreover, for each cluster, galaxies were first selected with \texttt{MAG\_AUTO\_I}$<m_{\rm lim}$, where 
\begin{equation}
m_{\rm lim} = -19.6 + 5\log_{10} \left[D_{c}(z)/(10 \, {\rm pc}) \right]\,,
\end{equation}
which is equivalent to a selection with \texttt{MAG\_AUTO\_I}$<23$ at $z=1$. This has been done in order to build an homogenous sample throughout the redshift range limited in rest-frame magnitude.\\
After using this galaxy membership assignment procedure, we then estimate, for each cluster, its fraction of red (and elliptical) galaxies by weighting each galaxy according to its membership probability. We compute the red (and elliptical) galaxy fraction for clusters containing at least five galaxies within $R = 2\,R_{500c}$. 





\begin{figure}
    \centering
    \includegraphics[width=0.5\textwidth]{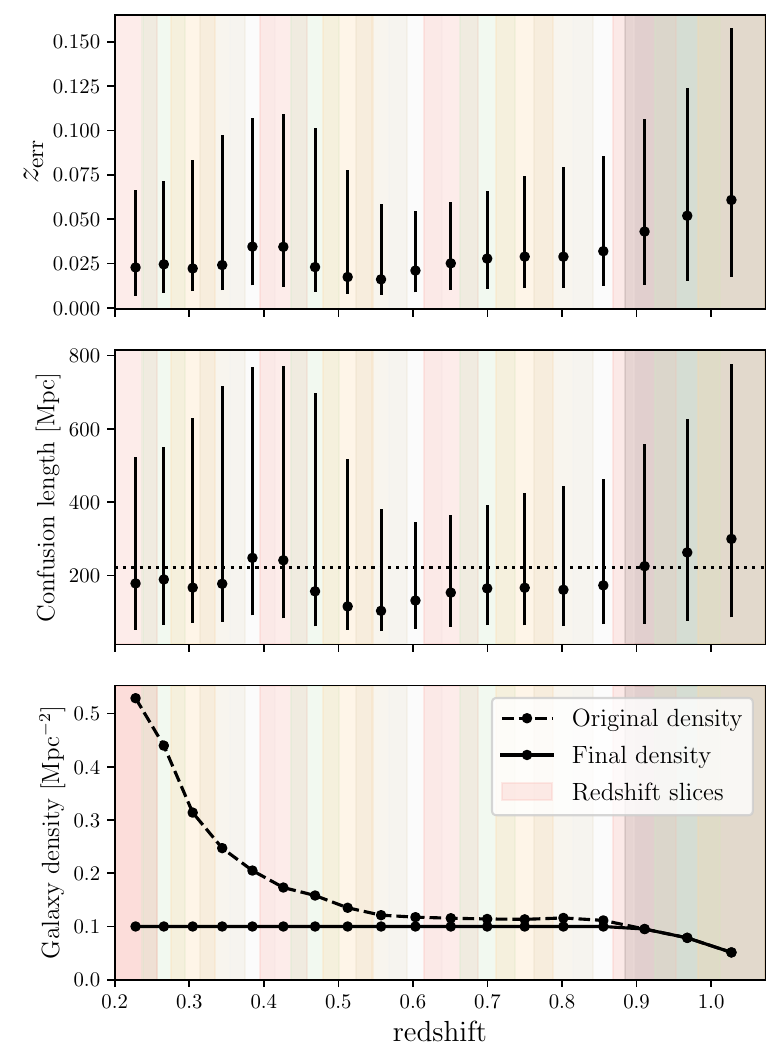}
    \caption{Photometric redshift uncertainty (top), corresponding confusion length (middle) and galaxy density per redshift slice (bottom) for our initial galaxy sample: the DESY3 \textsc{Maglim} catalog. The dashed line corresponds to the original effective galaxy density (see Table~\ref{tab:samples}) while the solid line is the final density (after sub-sampling) used for the cosmic web reconstruction to maintain a constant galaxy sampling through redshift. Each slice has a thickness of 222~cMpc (calibrated on the maximum confusion length over the redshift range). The span of each redshift slice is represented by color bands. The slices overlap with each other by $\sim 74$~cMpc. In the two upper panels the points correspond to the median of the distribution and the asymmetric errobars represent its 68\% quantiles. We discard the last three redshift slices (shaded) because of the decreased number density and increased confusion length.  }
    \label{fig:confusion_length}
\end{figure}

\begin{figure*}
\centering
    \includegraphics[width=0.95\textwidth]{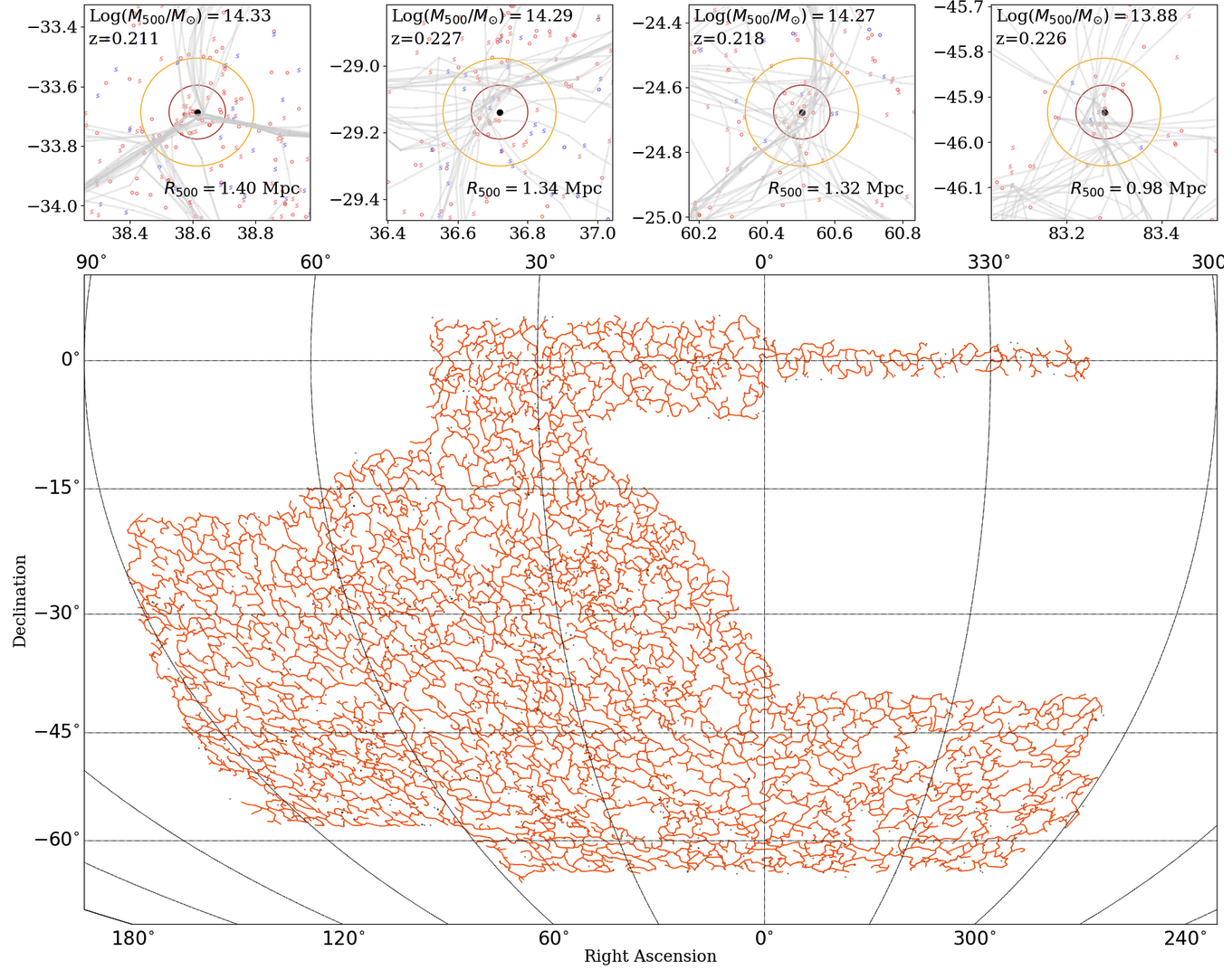} 
    \caption{Visualization of {\sc DisPerSE}  skeletons in the DES footprint, excluding masked areas for the first tomographic slice ($0.2<z<0.2562$). Top: Zoom-ins of a selection of four randomly chosen clusters, with Mass, redshift and $R_{500}$ as shown. Grey lines are the twenty random realizations, the black dot is the cluster center, the red circle marks $R_{500}$ and the orange circle marks 2$\times R_{500}$. Galaxies from the \citep{cheng2021} morphological catalog within the slice and with r-magnitudes brighter than r$<$25 are shown. Open circles mark elliptical galaxies and 'S' symbols mark spiral galaxies. Blue colours have (g-i)$<$1.3, and red colours have (g-i)$>$1.3.  Bottom: The skeletons of the first slice and realization are shown in orange.} 
    \label{fig:cl_vis}
\end{figure*}
\section{Cosmic web extraction}

\label{sec:cosmic_web}

\subsection{Tomographic redshift slices for 2D filament detection}
\label{sec:redshift_slices}

The confusion along the line-of-sight induced by the photometric redshift uncertainties makes a 3D reconstruction of the cosmic network impossible. Therefore, the reconstruction is performed in~2D in tomographic slices of fixed comoving thickness across the concerned redshift range \citep[see e.g.][]{Laigle2018,gouin2026,laigle2026}. The choice of slice thickness is informed by the typical photometric redshift uncertainties, as illustrated on Fig.~\ref{fig:confusion_length}. At $z\sim 0.9$, the typical photometric redshift uncertainty is $\sim 0.03$, which translates into a comoving confusion length along the line-of-sight of 
$\sim 220$~cMpc,  
(roughly corresponding to the interval encompassing 68\% of the redshift uncertainty around the redshift estimate $z_{\rm DNF}$). 
We therefore choose slices of comoving thickness $\sim 222$~cMpc with an overlap of $\sim 74$~cMpc. 
While all selected galaxies within the $222$~cMpc
-thickness are used for the reconstruction, the measurement will be perfomed only on galaxies and clusters associated to the central $\sim 74$~cMpc to avoid edge effects along the line-of-sight. 
In addition, to perform a consistent reconstruction of the cosmic web over the given redshift range, we require an homogeneous galaxy density in the different tomographic slices. In the \textsc{MagLim} catalogue, the galaxy density rapidly drops above $z>0.9$, as illustrated on Fig.~\ref{fig:confusion_length}. We therefore choose to perform the reconstruction over the redshift range $0<z<0.9$, with a galaxy density per slice of $\sim 0.1 {\rm gal}/{\rm Mpc}^2$. Galaxies only have a certain probability to be in a given tomographic slice, given their photometric redshift uncertainties. To incorporate this uncertainty in the reconstruction and to propagate it easily at later stages of the analysis, we perform 50~realisations of the catalog by randomly drawing redshifts under the redshift posterior of each galaxy, assuming the posterior to be gaussian with a standard deviation corresponding to the redshift uncertainty. To achieve a constant galaxy density across the redshift range, we sort galaxies that have been assigned to a tomographic slice based on their magnitude, and we select the number of galaxies required to match the target final density. {Extracting filaments at fixed number density of tracers ensures that the detected structures belong to the same physical scales, and thus, are comparable accross redshifts.} Table~\ref{tab:samples} provides a summary of the tomographic slices with the initial effective density.




\subsection{Cosmic filament extraction with {\sc DisPerSE} }


 The cosmic filaments are then extracted in 2D, on each of the 50 realizations of the galaxy distribution in the tomographic slices.
The {\sc DisPerSE}  \citep{Sousbie_Disperse_paper1, Sousbie2011b} description of the cosmic web is based on the analysis of the topology of the density field. The latter is computed from the distribution of galaxies using the Delaunay Tessellation Field Estimator
\citep[DFTE,][]{SchaapWeygaert2000, WeygaertSchaap2009}. Each tomographic slice is viewed as a 2D portion of the celestial sphere at the median redshift of the slice.  In this work, DTFE is applied in 2D on the stereographic projection of the sphere \citep[see a similar approach in][]{laigle2026}. Angles are preserved through the stereographic projection, and therefore the tesselation (i.e the way galaxies are connected through the triangulation) is the same on the sphere as on its projection. However, the projection does not preserve area. Therefore, in order to get a correct estimate of galaxy density (measured through the inverse area of the triangles) in the tomographic slice, we replace in the tesselation file the area of the projected triangles by the one it would have on the sphere. 
Using discrete Morse theory, {\sc DisPerSE}  identifies the critical points of this density field, which are are located where the gradient of the field vanishes. Filaments are then defined as the ridges in the Delaunay tessellation which connect the saddles of the density to the peaks. Each filament is defined as a set of small segments {whose sampling points are defined by the galaxy positions of the input catalogue}.

Importantly, the user can select the significance of the detected filaments with respect to Poisson noise by fixing the persistence ratio of the corresponding pair of critical points. In this work, we adopt a persistence ratio of 1.5$\sigma$, calibrated against simulated catalogues. This calibration is performed by computing the cross-correlation between the cluster-size halos and the critical points of the skeleton in different mass bins. We choose the persistence that optimizes the cross-correlation at all masses. The description of the simulation is presented in Appendix~\ref{sec:simdata} and the calibration is explained in Appendix~\ref{App:Fils_in_mocks}. 

\subsection{Connectivity measurement}
 For a given cluster associated with a given redshift slice, we measure connectivity in each of the 50 realisations.
Connectivity $\cal{C}$ of a cluster is defined as the number of filaments crossing a circle drawn around the cluster at $f\times R_{500}$, where $R_{500}$ is the radius inside which the mean density is 500 times the critical density of the Universe. We adopt $f=2$, but we do not notice significant difference in the results when varying $f$ around this value (from $f=1$ to 3). This definition of connectivity is commonly adopted in the literature \citep[e.g][]{darraghford19,boldrini24,gouin2026}, which is also referred to as multiplicity in some papers \cite{codis2018}. We adopt this definition because it corresponds to a physically measurable quantity in observations. Appendix~\ref{app:systematic effects} investigates eventual systematic effects on connectivity measurements due to the photometric redshift uncertainties and the presence of masks in the survey. In brief, photometric redshift uncertainties tend to disconnect the structures and to systematically lower the connectivity by about 1. This was already noted by \cite{darraghford19}. The presence of masks in the survey has no effect on average, since they cover a very small effective area.




\section{Results}
\label{sec:results}

An example of the cosmic filaments extracted by {\sc DisPerSE} in a tomographic slice is shown in Fig.~\ref{fig:cl_vis}. The bottom portion of the figure shows all of the filaments extracted in the range $0.200<z<0.2562$, excluding masked areas, for one realization. At the top of the figure, we zoom-in around four randomly chosen clusters. Here, we present 50 random realizations of the filaments around the clusters. Many filaments are seen to cross into the cluster core. Many of the realizations overlap, showcasing the structures with the highest fidelity. 

\subsection{Mass-connectivity relation}



\begin{figure*}
    \centering  
    \includegraphics[width=0.95\linewidth]{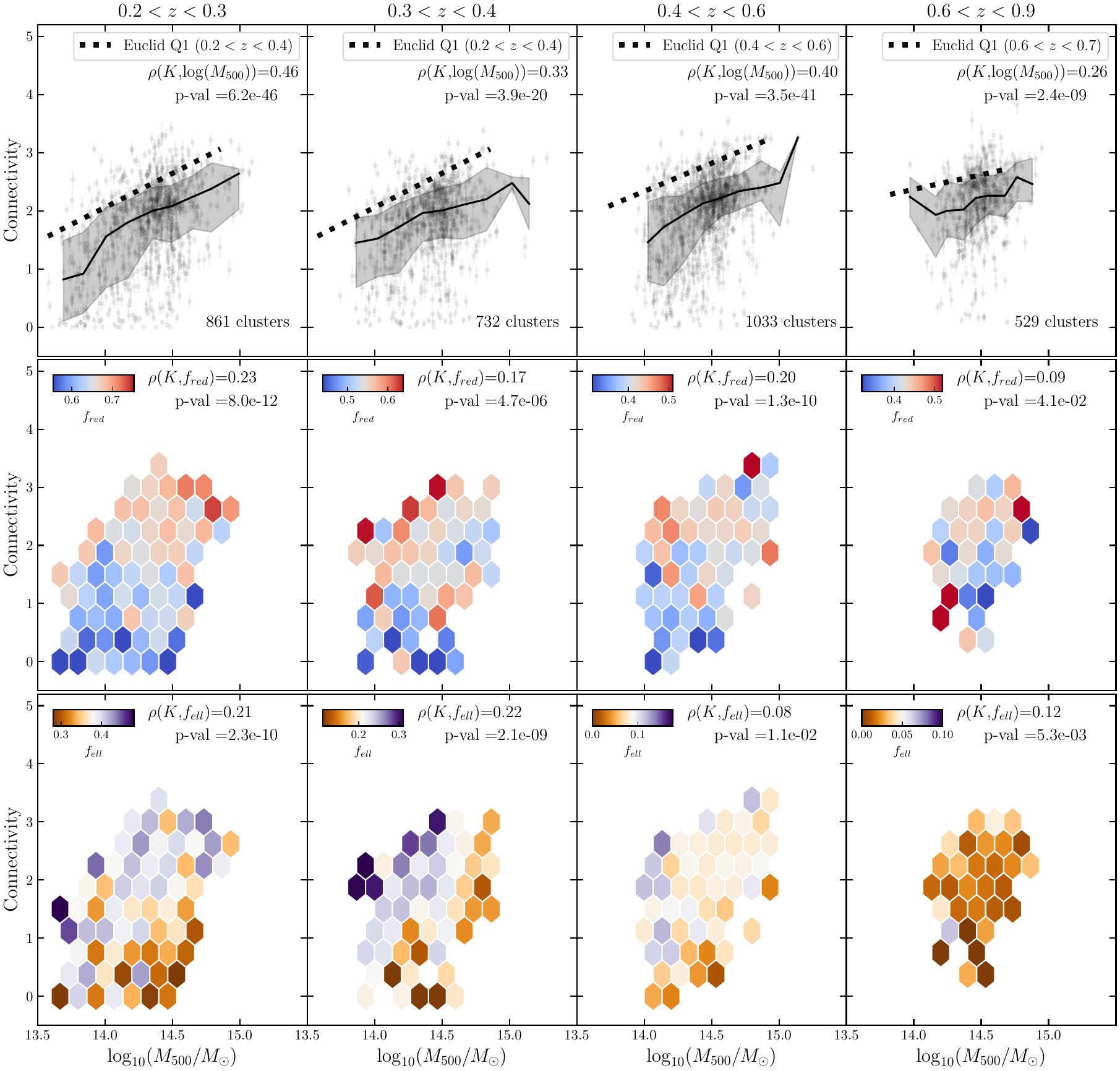}
    \caption{
    Top: Mean relation between connectivity and cluster mass in four redshift bins. The shaded area indicates the dispersion around the mean. In each panel, the Pearson correlation coefficient $\rho$ and its associated p-value are indicated. The black dashed lines show the fitted mean relation from \cite{gouin2026} in comparable redshift bins.
    Middle and bottom: Mean relation between connectivity and cluster mass in four redshift bins, colour-coded by the fraction of red galaxies $f_{red}$ (middle) and elliptical galaxies $f_{ell}$ (bottom). In each panel, the partial Pearson correlation coefficient $\rho$ (after accounting for the dependence on cluster mass) and its associated p-value are indicated.
    \label{fig:connectivitymass}}
\end{figure*}

The top panels of Fig.~\ref{fig:connectivitymass} present the relation between connectivity $K$ and cluster mass $\log(M_{500c} / M_{\odot})$ in 4 different redshift bins.
Each data-point corresponds to a given cluster, with the connectivity being the mean over 50 realisations, and the errorbars being the standard error on the mean connectivity. 
Overall, we recover the trend that connectivity increases with cluster mass, with a positive value of the Pearson correlation coefficient. In the highest redshift bin, the number of low-mass clusters is small, which limits the mass range over which the correlation can be computed, and therefore the significance of the correlation. 

For comparison with recent observational results, the red lines show the mean connectivity–mass relation measured from Euclid Q1 data in comparable redshift bins \citep[see Fig.8][]{gouin2026}. That analysis employed a similar methodology, using a 2D slice centred on the cluster redshift, filament extraction with {\sc DisPerSE} at a persistence threshold of 1.5 (similar to our analysis), and connectivity measured within $2\times R_{500c}$. Overall, we find good agreement with these measurements, although our connectivity values are systematically lower across all masses, and especially at $z<0.6$. We note that the calibration of the reconstruction as we carried out in Appendix~\ref{App:Fils_in_mocks} was not performed in \citet{gouin2026}. Therefore, the systematic lower connectivity for a similar parametrisation of the reconstruction could be driven by different selection of the galaxies used to reconstruct the filaments, leading to a lower galaxy density in the DES Y3 dataset. In particular, the \textit{Euclid} sample used in \citet{gouin2026} is mass-selected while the DES Y3 dataset used in the present paper is magnitude-selected, because stellar mass information is unavailable. Furthermore, in the present study, we selected galaxies to maintain a close-to-constant galaxy density through redshift, which lowers the galaxy density in slices at $z<0.6$. Since this selection was not done in \citet{gouin2026}, their connectivity might be slightly larger at low redshift for a similar parametrisation.

\subsection{Relation between connectivity and galaxy member properties}

Beyond its relevance for cosmology, the connectivity of galaxy clusters has been found to correlate with galaxy member properties \citep{gouin2026}. 
To interpret this result, the authors proposed that clusters of same mass but different connectivities have different formation pathways, which could also have impacted the properties of galaxy populations that they have accumulated \citep[see also][]{darraghford19}. In order to confirm or {deny} this interpretation, we present below the relation between connectivity and galaxy member properties, beyond the first-order mass effect.

The middle and bottom lines of Fig.~\ref{fig:connectivitymass} present the relation between connectivity and cluster mass colour-coded by the properties of galaxy cluster members.
More specifically, we first measure for each cluster the fraction of red galaxies $f_{\rm red}$ within $2\times R_{500}$. Each hexagon is colour-coded by the average value of $f_{\rm red}$ for all clusters falling into this mass and connectivity bin. We separate the measurements in different redshift bins to account for the fact that $f_{\rm red}$ also evolves with redshift. We then compute the partial Pearson correlation coefficient between connectivity $K$ and red fraction $f_{\rm red}$, beyond the first-order cluster mass dependency. Whatever the redshift, we obtain a positive correlation between the fraction of red galaxies and the connectivity, beyond the cluster mass dependency. The correlation is non-negligible and significant for the three lowest redshift bins, with partial Pearson correlation coefficients of $\rho = 0.23$ (p-value $8.0\times 10^{-12}$), $\rho = 0.17$ (p-value $4.6\times 10^{-6}$), $\rho = 0.20$ (p-value $1.3\times 10^{-10}$) in the redshift bins $0.2<z<0.3$, $0.3<z<0.4$, and $0.4<z<0.6$ respectively. 
We then repeat the analysis with the fraction of elliptical galaxies $f_{\rm ell}$ per cluster. A similar result is achieved, with the galaxy population in more connected clusters being more elliptical than in clusters with low connectivity, at a fixed mass.  In this case, the correlation is significant for the two lowest redshift bins, with partial Pearson correlation coefficients between connectivity $K$ and elliptical fraction $f_{\rm ell}$ of $\rho = 0.21$ (p-value $2.3\times 10^{-10}$), $\rho = 0.22$ (p-value $2.1\times 10^{-9}$), in the redshift bins $0.2<z<0.3$ and $0.3<z<0.4$, respectively. We simply note that the correlation between connectivity and elliptical fraction drops more rapidly in the bin $0.4<z<0.6$  than between connectivity and red fraction. 
Interestingly, applying a stricter criterion for the reliability of morphological classifications (i.e. \texttt{confidence\_flag}$>1$ instead of $>0$) strengthens the correlation between connectivity and elliptical fraction in the redshift bin $0.4<z<0.6$, increasing the partial Pearson correlation coefficient to $\rho=0.14$ (p-value $=2.4\times10^{-5}$). However, this stricter selection reduces the number of cluster member galaxies, making the correlations less well constrained at lower redshifts.

For completeness, corresponding maps displaying the standard errors on the measurements, as well as the  p-value associated on the partial Pearson correlation coefficient are presented in appendix~\ref{App:noise_map}.

These results could be interpreted in a framework where more connected clusters have grown through more frequent cluster major mergers and interactions. As a matter of fact, the merging of two groups or clusters will increase not only the mass of the remnant, but also its connectivity (provided that at least one of the progenitor is connected to more than 2 filaments), each of the progenitor bringing in  its connected filaments \citep[see e.g. the discussion in][]{darraghford19}. It is reasonable to think that frequent interactions at the cluster scale will also impact galaxies within it, or in other words it is easier to preserve a large fraction of spiral and star-forming galaxies when the cluster smoothly accretes matter from its own filaments. Furthermore, we could take the connectivity as a proxy of the large-scale overdensity around clusters. Galaxies in clusters with high connectivity, hence denser environments, are more susceptible to being pre-processed. In this case, the resulting accreted population in the clusters would contain more elliptical and red galaxies than in same-mass clusters of lower connectivity.

\subsection{Galaxy pre-processing in filaments around clusters}

\begin{figure*}[!h]
    \centering
        \includegraphics[width=0.95\linewidth]{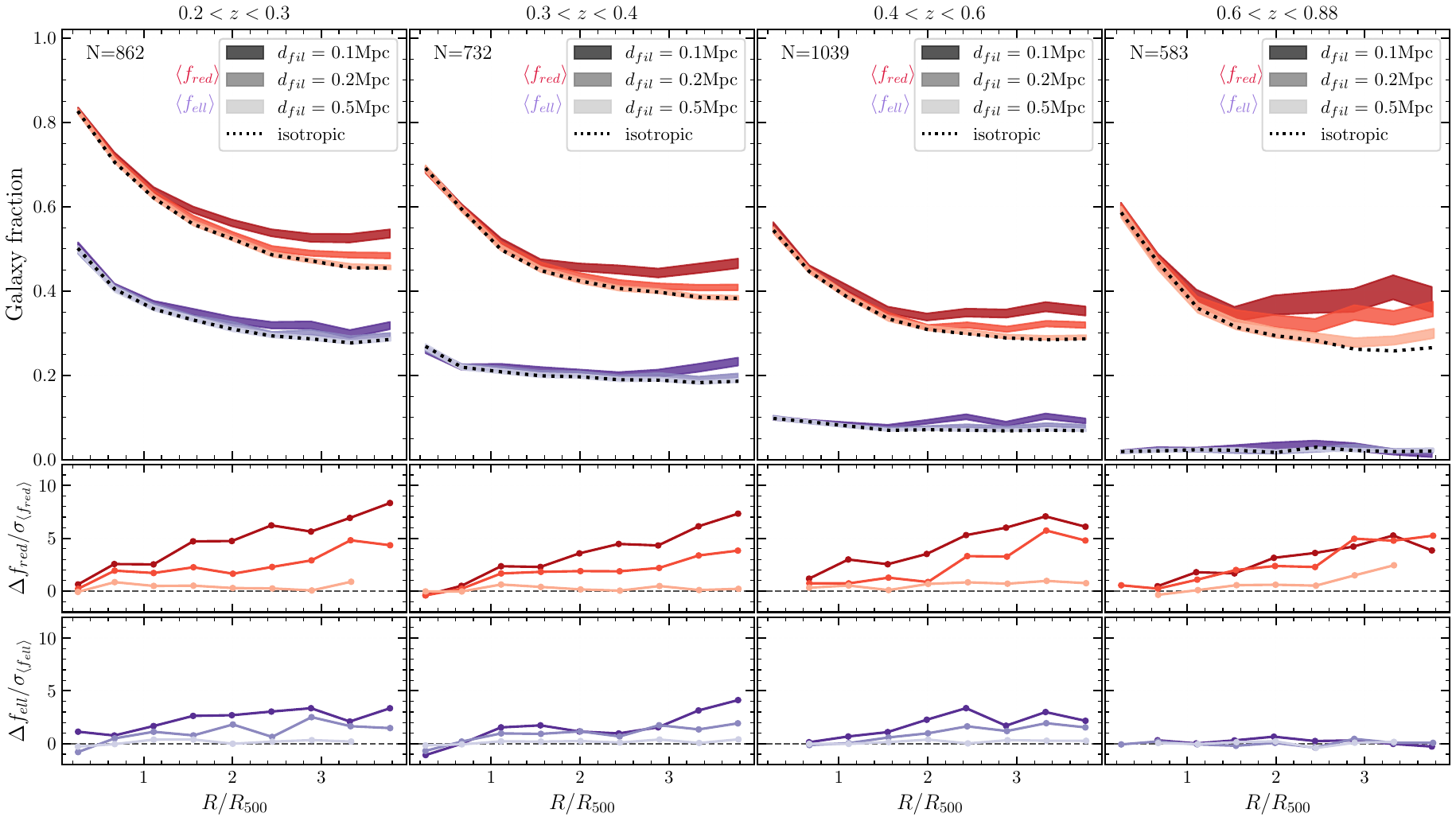}
    \caption{Mean radial profiles of the elliptical (purple) and red (red) galaxy fractions around clusters, shown for isotropic (dotted lines) and anisotropic (solid lines) measurements across the four redshift bins (one panel each). Anisotropic profiles are computed by selecting galaxies within different maximum distances to filaments ($0.1$, $0.2$, and $0.5,\mathrm{Mpc}$), as indicated by the color code. Line thickness reflects the uncertainty on the mean profile, estimated over the full cluster sample.}
    \label{fig:red_elliptical_fraction_profile}
\end{figure*}

We now turn to investigating the role of galaxy pre-processing  within filaments. To do so, the top panels of Fig.~\ref{fig:red_elliptical_fraction_profile} present the fractions of elliptical and red galaxies, $f_{\rm elliptic}$ and $f_{\rm red}$ in concentric annuli centred on galaxy clusters in the four different redshift bins.   We compare isotropic profiles, computed using all galaxies within each annulus, with anisotropic profiles, obtained by restricting the analysis to galaxies located within a maximum distance $d_{\rm fil}$ from the filament spines. The radial profiles are averaged over the cluster sample, with uncertainties estimated from the standard error on the mean.
The bottom panels present the relative difference between the anisotropic and the isotropic profiles (normalized by the square root of the quadratic sum of the uncertainties on each profile).

For large filament distances ($d_{\rm fil}=0.5\,{\rm Mpc}$), the isotropic and anisotropic profiles are nearly indistinguishable. At such distances, the selected galaxy population in the anisotropic case encompasses most of the galaxies within the annuli, resulting in profiles that closely resemble the circular average. As the filament-distance threshold decreases, however, the discrepancy between isotropic and anisotropic profiles becomes increasingly pronounced. We do not consider distances smaller than $0.1\,{\rm Mpc}$ in order to retain a sufficient large number of galaxies within the filaments, for robust estimates of the red and elliptical fractions in each radial bin. For $d_{\rm fil}=0.1\,{\rm Mpc}$, we observe a significant difference between the two measurements: galaxies located close to filaments exhibit systematically higher red and elliptical fractions than those inferred from the isotropic average. This trend indicates that galaxies embedded in filaments around clusters are, on average, more evolved than the overall cluster-centric galaxy population at the same radius. The trend is much more pronounced for the red fraction than for the elliptical fraction, suggesting that pre-processing is more a matter of turning galaxies red than transforming them in ellipticals.

These results suggest a tendency toward a filament-driven pre-processing scenario, whereby galaxies experience 
star-formation transformations prior to cluster infall. This interpretation is consistent with previous observational studies reporting an excess of passive, red galaxies preferentially distributed along filaments \citep{Martinez2016,Sarron2019,Gouin2020,Stephenson2025}. We note that the fact that pre-processing around clusters seems to affect more the blue-to-red transformation than the morphological transformation is consistent with a growing literature on this topic \citep[e.g.][]{ECgentile2026}. However, full consensus has not been reached, for example \citep[e.g.][]{Igel26} find large scale pre-processing only to be significant at $z>0.5$. 
Such discrepant results could be due to different biases owing to selection effects. Consistent extraction of the cosmic web around clusters over larger volumes and redshift ranges are needed to reach an agreement on this question.
We also examined whether the strength of filament-driven pre-processing varies with cluster connectivity. However, the results remain inconclusive, preventing us from establishing a clear relation between the two.

The physical origin of filament-driven pre-processing remains debated, with several competing mechanisms potentially at play. A first possibility is that the observed signal is not driven by filaments themselves, but by the population of galaxy groups embedded within filaments. In this picture, filaments primarily act as dynamical channels funnelling pre-processed group galaxies into clusters. Indeed, \citet{Kuchner2022}, using the \texttt{TheThreeHundred} simulation suite, provided an inventory of galaxies accreted by clusters through filaments, showing that up to 45\% of infalling galaxies enter clusters via filamentary structures, and that approximately 12\% of these filament-accreted galaxies belong to bound groups. This supports a scenario in which a significant fraction of pre-processing occurs in group environments prior to cluster infall, with filaments serving mainly as transport pathways.
Beyond group-driven pre-processing, filaments may also directly influence galaxy evolution. Enhanced merger rates within filamentary environments could accelerate stellar mass assembly and induce morphological transformation, leading to earlier quenching, as recently demonstrated by \citet{Zakharova2026} using the TNG100 simulation. In addition, gas dynamics within filaments may also play a direct role. Processes such as cosmic web detachment, in which cold gas accretion streams that fuel galaxies may be disrupted while traversing cosmic filamentary structures, have been proposed as a mechanism to suppress star formation \citep{Calvo2019}. Alternatively, filament-specific dynamical effects, such as enhanced vorticity and shear flows, could contribute to stripping or heating of gas reservoirs, thereby promoting quenching \citep{song21}.
Disentangling these scenarios remains challenging, as they are not mutually exclusive and may operate concurrently. Future work combining higher-resolution simulations and deeper observational data will be required to separate the relative contributions of group pre-processing, filament-induced merger activity, and hydrodynamical effects within the cosmic web.











\section{Conclusion}
\label{sec:conclusion}

Our study measured the cosmic web connectivity of galaxy clusters, using the DES Y3 dataset to extract cosmic web filaments. The cosmic web extraction was calibrated to optimize filament detection around halos more massive than galaxy groups. To account for photometric redshift uncertainties, we generated 50 realizations of the cosmic skeleton through a Monte Carlo resampling of individual galaxy redshift probability distribution functions. We then counted, in each realization, the number of filaments branching out of clusters detected in X-ray and SZ, thereby measuring a quantity referred to as connectivity. Our results are as follows:

\begin{itemize}

    \item We measured the mass–connectivity relation in four different redshift bins. This relation is consistent with the literature and shows no significant redshift evolution within our current statistical uncertainties;

    \item After selecting cluster member galaxies within $2R_{500}$, we quantified the correlation between connectivity and secondary cluster properties, namely the red and elliptical fractions of member galaxies. After accounting for cluster mass, we find a significant positive correlation, indicating that more connected clusters tend to host a higher fraction of red and elliptical galaxies;

    \item By probing radial profiles of elliptical and red galaxy fractions around clusters (up to $4R_{500}$), we observe a marginally significant enhancement along filament directions. Anisotropic radial profiles show a larger excess of red than elliptical galaxies in regions around filaments, suggesting that pre-processing is more efficient at quenching galaxies than transforming their morphology in cluster surroundings.

\end{itemize}


To conclude, we emphasize that the main source of uncertainty in these findings arises from projection effects and the use of photometric redshifts. Improving photometric redshift accuracy would allow the use of thinner tomographic slices, thereby mitigating projection effects and the blurring induced by redshift uncertainties. Increasing the sample size would reduce statistical uncertainties. Both improvements will be achieved with the next releases of the \textit{Euclid} survey \citep{mellier2025} and the Vera C. Rubin Observatory Legacy Survey of Space and Time (Rubin-LSST \citep{Ivezic2019}. Both of which will therefore provide significantly stronger constraints on cosmology using connectivity metrics (e.g., the amplitude and slope of the connectivity–mass relation). Despite the limitations imposed by photometric redshift uncertainties and projection effects, our reconstruction reliably traces large-scale cosmic filaments, although it is not designed to resolve smaller-scale structures. These filaments provide promising targets for future weak-lensing stacking analyses using the DES Y3 and DECADE mass maps \citep{Gatti2026}.
\begin{acknowledgements}
This research was supported in part by the National Science Foundation under Grants No. NSF PHY-1748958 and PHY-2309135 to the Kavli Institute for Theoretical Physics (KITP).
We would like to thank Giulia Giannini and Joe DeRose for providing the \textsc{MagLim} selection on the \textsc{Buzzard} simulations, Florian Sarron and Tony Bonnaire for useful discussions.
 PGV is supported by MIUR/PRIN 2022 "Optimal cosmology with next generation galaxy surveys". LOVE was supported by NSF grant AST-2205976 which also supported work by GG. This material is based in part upon work while LOVE served at the U.S. National Science Foundation. Any opinions, findings, and conclusions or recommendations are solely those of the authors and do not necessarily reflect the views of the U.S. NSF or Federal government.
\end{acknowledgements}
\bibliographystyle{aa} 
\bibliography{bibliography}

\appendix

\section{Simulated data}
\label{sec:simdata}

Our methodology is calibrated and validated using simulated catalogs. In particular, we use one realisation of the \textsc{Buzzard} $N$-body simulation suite \citep{DeRose2022}. \textsc{Buzzard} produces synthetic galaxy catalogs built starting from
$N$-body lightcones generated by \textsc{L-GADGET2} \citep{Springel2005}, and spans
$10313\, {\rm deg}^2$ out to $z = 2.35$. It assumes a flat $\Lambda$CDM cosmology with $\Omega_m = 0.286$, $\sigma_8 = 0.82$, $\Omega_b= 0.047$, $n_s= 0.96$, $h = 0.7$. The simulations are created starting from three lightcones with different
resolutions and sizes ($1050^3$, $2600^3$ and $4000^3$ Mpc$^3 \, h^{-3}$ boxes and
$1400^3$, $2048^3$ and $2048^3$ particles), since at high redshift the box needs to be larger. Halos are identified using the public code
\textsc{ROCKSTAR} \citep{Behroozi2013} and they are populated with
galaxies using the \textsc{ADDGALS} algorithm \citep{DeRose2019}, which provides
positions, velocities, magnitudes, spectral energy distributions (SED) and ellipticities.  Galaxies are assigned their properties based on the relation between redshift, $r$-band absolute magnitude, and large-scale density from a sub-halo
abundance matching model \citep{Conroy2006, Lehmann2017} in higher resolution $N$-body simulations. SEDs are assigned to galaxies by imposing the matching with the SED-luminosity-density relationship measured in the SDSS data. SEDs are $K$-corrected and
integrated over the DES filter bands to generate DES $grizY$ magnitudes. In order to reproduce the \textsc{MagLim} sample itself in the simulation, the \textsc{DNF} code has been run on a subset of one
\textsc{Buzzard} realization. We use the same  \textsc{MagLim} \textsc{Buzzard} catalog used in \citet{Porredon2022} and \citet{Giannini2022}.  For further details regarding the \textsc{Buzzard} simulation specifications and validation of relevant observables see Sections~3.1 and~3.2 in \citet{DeRose2022} and Section~2.8.1 from \citet{Giannini2022} .

\section{Calibration of the persistence for filament extraction}
\label{App:Fils_in_mocks}

\begin{figure*}[!h]
    \centering
    \includegraphics[width=0.95\textwidth]{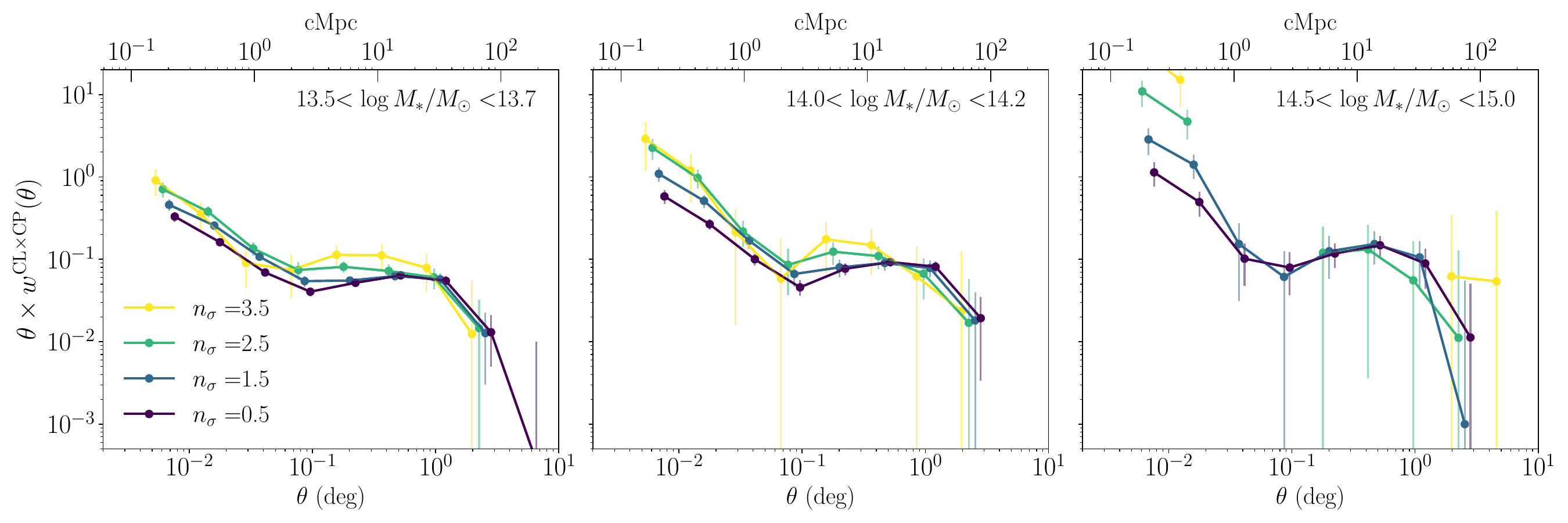}    
    \caption{Cross-correlation $\theta \times w_{{\rm CL}\times{\rm CP} }$ between halos in different mass bins (from left to right) and maxima (CP) in skeletons extracted in different persistences (purple to yellow curves). The cross-correlation is averaged here in 3 tomographic slices.  We display $\theta w(\theta)$ to decrease the dynamical range. }
    \label{fig:crosscorrel}
\end{figure*}
We use the \textsc{Buzzard} simulated catalogues described in App.~\ref{sec:simdata} to calibrate the parametrization of the filament finder. 
Since our goal is to measure cluster connectivity, we aim to select a persistence threshold that best reproduces the association between massive halos and the maxima of the filamentary skeleton.

The content of the filament catalogues extracted with {\sc DisPerSE} depends on the persistence threshold, i.e. the ratio of the field value at the position of the two critical points of a pair (saddle - peak). Expressed in number of $\sigma$, persistence quantifies the significance of the pair  with respect to Poisson fluctuations. {\sc DisPerSE} offers the option to keep in the catalogue only those filaments connecting critical point pairs with persistence higher than a chosen threshold. Persistence is therefore a way to filter only robust filaments with respect to the noise. The optimal choice for persistence depends on the quality of the sampling of the density field (i.e. the density of tracers and their probability to be on the slice).  

In this work, we focus on  large-scale filaments connecting to these clusters, and therefore we calibrate the persistence so that these filaments are correctly extracted. Smaller scale filaments obviously exist, but are more likely to be buried in noise due to the large slice thickness and high redshift uncertainties of the galaxies tracing the density field. Inspired by a method designed in \citet{GalarragaEspinosa2024}, 
we choose to optimize the calibration so that massive halos fall as much as possible at nodes of the cosmic web (i.e being one point of the critical pairs defining filaments). To do so, we measure the projected angular cross-correlation between halos in various mass bins and the maxima of the skeleton. We want to minimize the signal at large separation


To perform this calibration, we generate {\sc DisPerSE} skeletons from the simulated catalogue (with photometric redshifts uncertainties comparable to the data) at persistence $n_{\sigma}=0.5$, $1.5$, $2.5$, $3.5$. For each extraction and each tomographic slice, we compute the projected angular cross-correlation between halos in various mass bins and the maxima of the skeleton, using {\tt Treecorr} \citep{jarvis2004}. Error bars are obtained from jackknifing subregions (7~patches used\footnote{Varying the number of patches does not significantly impact the estimation of error bars.}). A random catalogue is generated with 500~000 points, with the same geometry as the DES footprint. 

Fig.~\ref{fig:crosscorrel} displays the inverse variance-weighted mean and error across 20 different realisations in three tomographic slices. We choose to plot $\theta w(\theta)$ to decrease the dynamical range. Overall, the cross-correlation signal increases with increasing halo mass, as expected. The small-scale correlation signal ($\theta <0.1$~degree) is generally higher for more persistent skeletons, but we note a significant signal at intermediate scale (secondary bump at $\theta \sim 0.5 - 1$~degree), indicating that a significant number of points in one dataset do not have a close counterpart in the other datasets, and are associated with a more distant one. At low halo mass $13.5 < \log M_*/M_\odot < 13.7$, this secondary bump is more significant in high persistence skeletons ($n_{\sigma} = 2.5$ and $n_{\sigma} = 3.5$, which have a low-number of critical points, all corresponding to the highest overdensities), indicating that many low-mass halos do not find a close counterpart maximum in the skeleton when the persistence is too high. At high halo mass  $14.5 < \log M_*/M_\odot < 15.0$, this signal is larger for $n_{\sigma} = 0.5$, indicating that many maxima of the skeleton do not have a close counterpart in the halo catalog if the persistence is too low. In addition, we note that at high halo mass the number of close pairs is very sparse in skeletons with $n_{\sigma} > 1.5$, leading to a sporadic signal (only a handful of pairs are found). For this reason, we choose to set $n_{\sigma} = 1.5$ in our analysis. 

\section{Systematic effects on filament extraction}
\label{app:systematic effects}
\begin{figure}
    \centering
    \includegraphics[width=0.5\textwidth]{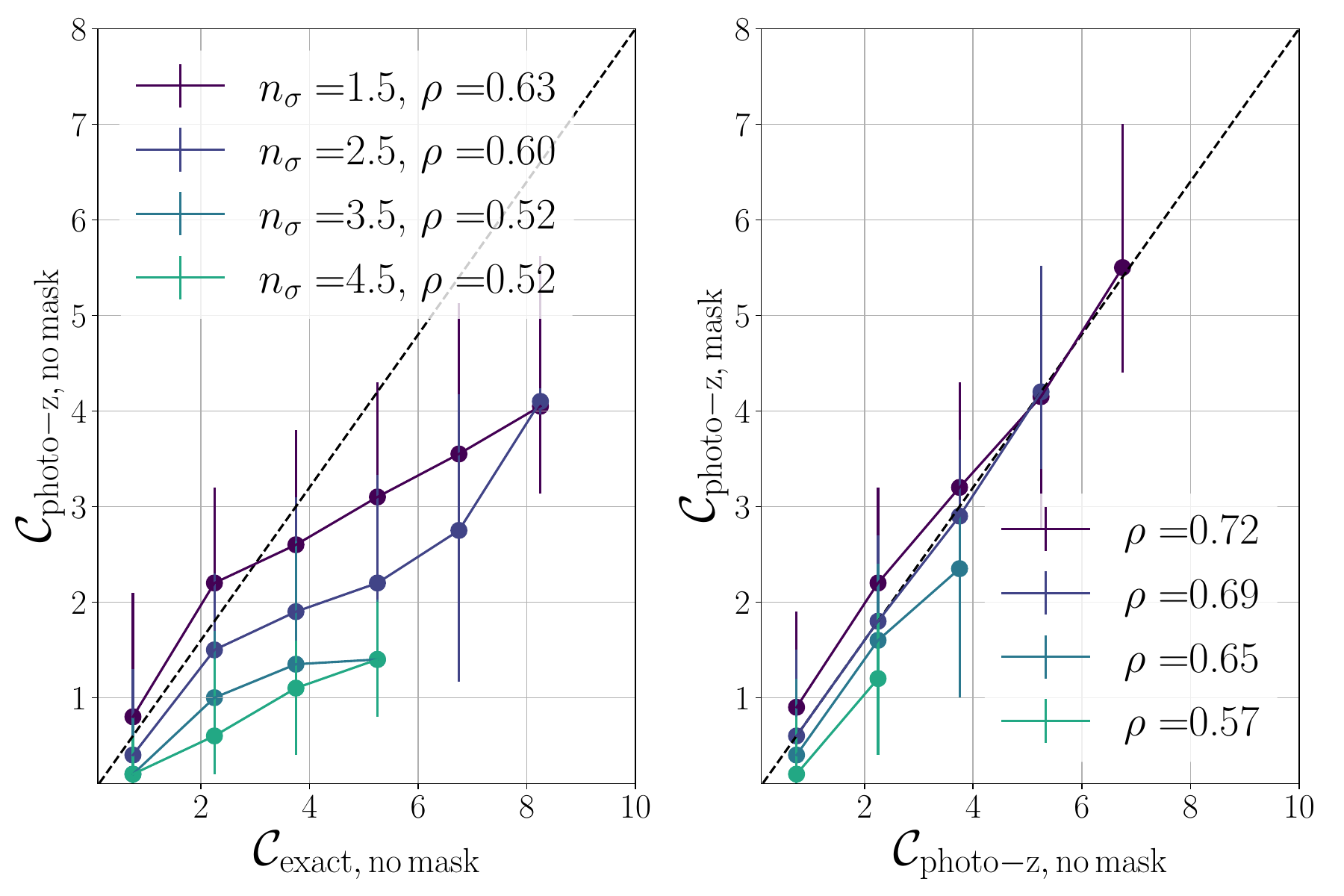}
    \caption{\textit{Left:} Correlation between the connectivity of halos when the skeleton is computed with the exact redshift ($x$-axis) or with the photo-$z$ ($y$-axis). The Pearson correlation coefficient is $\rho = 0.63$ when the persistence is $n_{\sigma} = 1.5$ (persistence calibrated in Appendix~\ref{App:Fils_in_mocks}). \textit{Right:} Correlation between the connectivity of halos when the skeleton is computed without ($x$-axis) or with ($y$-axis) survey masks.}
    \label{fig:impact_err_connect}
\end{figure}

Relying on the \textsc{Buzzard} simulation described in App.~\ref{sec:simdata}, we test the impact of photo-$z$ and survey masks on the skeleton reconstruction. Specifically, we extract the skeleton on a version of the catalogue which does not include photo-$z$ uncertainties, and also on a version of the catalogue which does not contain survey mask. 
Fig.~\ref{fig:impact_err_connect} presents the impact of these two effects on connectivity. In the left panel, we measure the correlation between the connectivity of halos when the skeleton is computed with the exact redshift ($x$-axis) or using the photo-$z$ ($y$-axis) for different persistence thresholds. As was noted before in \cite{darraghford19}, photometric redshift errors tend to disconnect the density field, hence lower the connectivity, but the Pearson correlation coefficient remains high ($\rho = 0.63$ when the chosen persistence is the one we calibrated in Appendix~\ref{App:Fils_in_mocks}). The right panel presents the correlation between the connectivity when the skeleton is computed with and without survey masks.
Survey masks  have very little impact on connectivity. We notice a slight decrease of connectivity in the high persistence skeleton when survey masks are included.

\section{The dispersion of cluster properties on connectivity-mass diagram \label{App:noise_map}}

\begin{figure*}
    \centering  
    \includegraphics[width=0.95\linewidth]{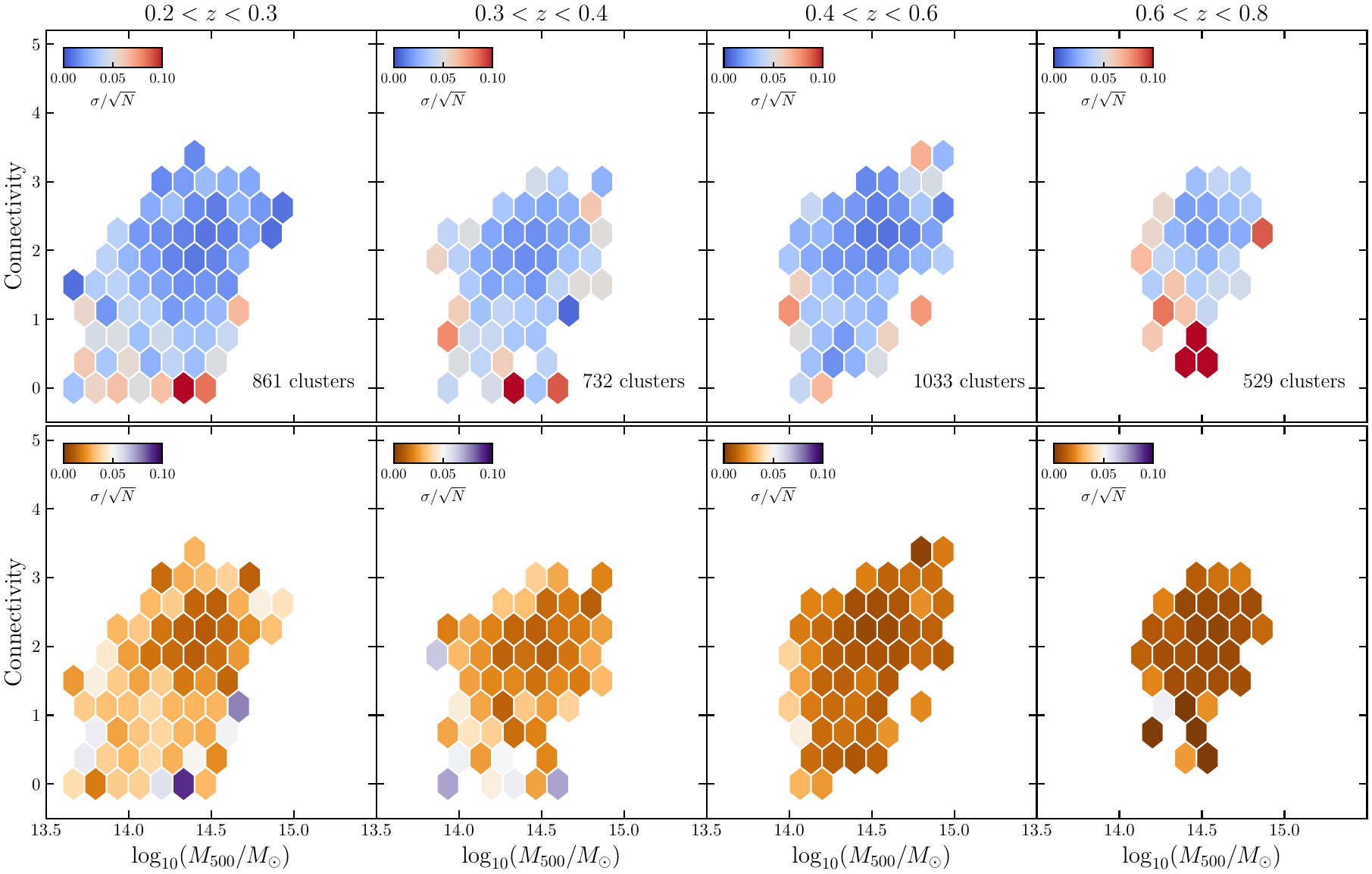}
    \caption{Same as Fig.~\ref{fig:connectivitymass} with the standard deviation.
    \label{fig:connectivityERR}}
\end{figure*}

In addition to the analyses carried out in Sect.~\ref{sec:results}, we show for completeness standard error on cluster properties on the connectivity-mass diagram in Fig.~\ref{fig:connectivityERR}. 

\section{Segregation between passive and star-forming galaxies}
\label{app:segreg}

\begin{figure}
\centering
    \includegraphics[width=0.98\columnwidth]{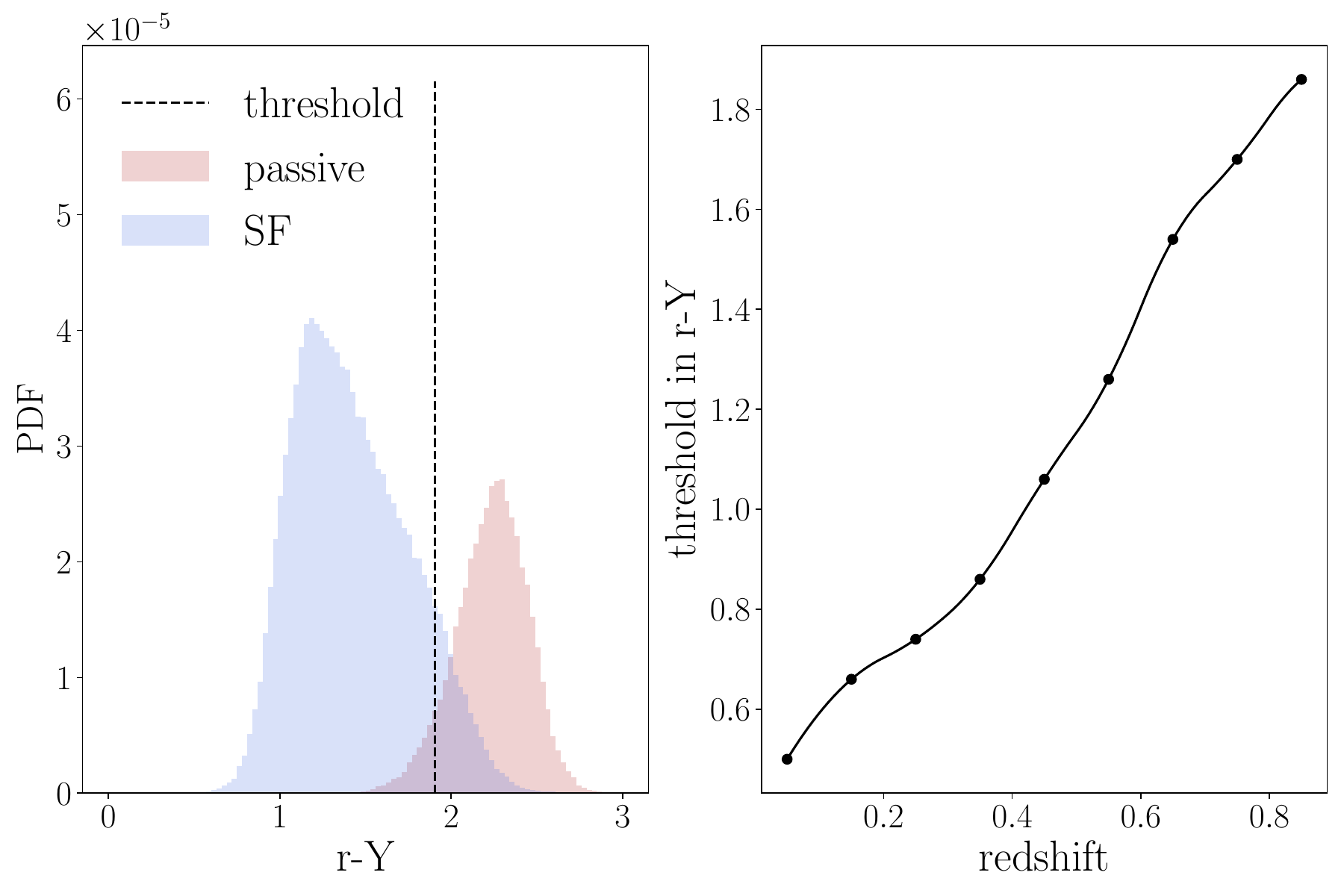} 
    \caption{Left: Calibration of the separation between red and blue galaxy populations, which we designed so that they match the best the passive and star-forming galaxy populations. The left panel displays the $r-Y$ distribution of passive and star-forming galaxies from COSMOS2020, the separation of which is based on the ${\rm NUV}rJ$ diagram. At each redshift, we determine the threshold in $r-Y$ which best separate both populations. The evolution of this threshold as a function of redshift is shown in the right panel. This threshold is then applied to the DES galaxy population.}
    \label{fig:threshold}
\end{figure}

The separation between star-forming and passive galaxies is estimated based on the $r-Y$ apparent color, and empirically calibrated on the COSMOS2020 catalogue \citep{Weaver2020}. More specifically, star-forming and passive galaxies are segregated in the COSMOS2020 catalogue based on the ${\rm NUV}rJ$ rest-frame colour diagram \citep{Weaver2020}. Then we estimate the normalized $r-Y$ apparent color distribution of star-forming and passive populations separately, and we determine in narrow redshift bins the  $r-Y$  threshold corresponding to minimum overlap between both populations. Here the $r$ and $Y$ magnitudes are taken fro  the second public data release of the Hyper Suprime-Cam (HSC) Subaru Strategic Program\footnote{We note that the filter response of HSC and DECAM in the $r$ and $Y$ filter passbands are very similar.} \citep{2019PASJ...71..114A}. The approach is illustrated in Fig.~\ref{fig:threshold}. We adopt this threshold to separate blue and red galaxies in our DES sample, according to their  $r-Y$ apparent colors and their redshifts.

\end{document}